\documentclass[preprintnumbers,amsmath,amssymb,floatfix]{revtex4}

\usepackage{graphicx}
\usepackage{dcolumn}  
\usepackage{bm}  

\begin{document}
\setlength\arraycolsep{2pt} 
\def\brho{{\mbox{\boldsymbol $\rho $}}}
\def\bomega{{\mbox{\boldsymbol $\Omega $}}}
\def\bomicron{{\mbox{\boldsymbol $\omega $}}}
\def\bnabla{{\mbox{\boldsymbol $\nabla $}}}
\def\ul#1#2{\textstyle{\frac#1#2}}
\newcommand {\vct}[1] {\mathbf {#1}}

\title{Quenched Charge Disorder and Coulomb Interactions }
 
\author{Ali Naji}
\email{naji@ph.tum.de}
\affiliation{Physics Department, Technical University of Munich, James Franck St., 
                  D-85748 Garching, Germany. 
                  }

\author{Rudolf Podgornik}
\email{rudolf.podgornik@fmf.uni-lj.si}

\affiliation{Faculty of Mathematics and Physics, 
University of Ljubljana,  SI-1000 Ljubljana, Slovenia and 
Department of Theoretical Physics, J. Stefan Institute, SI-1000 Ljubljana, Slovenia}
                 
\date{July 2005}

\begin{abstract}
We develop a general formalism to investigate the effect of quenched fixed charge disorder on 
effective electrostatic interactions between charged surfaces in a one-component 
(counterion-only) Coulomb fluid. Analytical results are explicitly derived for two asymptotic and complementary 
cases: i) mean-field or Poisson-Boltzmann limit (including Gaussian-fluctuations correction),
 which is valid for small electrostatic
coupling, and ii) strong-coupling limit, where electrostatic correlations mediated by counterions
become significantly large as, for instance, realized in systems with high-valency counterions. 
In the particular case of  two apposed and ideally polarizable planar surfaces with equal mean surface charge, we
find that  the effect of the disorder is nil on the mean-field level and thus the plates repel. In
the strong-coupling limit, however, the effect of charge disorder turns out to be additive in
the free energy and leads to an enhanced long-range attraction between the two surfaces. 
We show that the equilibrium inter-plate distance between the surfaces decreases for elevated disorder strength 
(i.e. for increasing mean-square deviation around the mean surface charge),  and eventually tends to zero, 
suggesting a disorder-driven collapse transition. 
\end{abstract}


\maketitle

 \section{Introduction}

 Electrostatic interactions usually provide the repulsion that stabilizies charged colloids and are one of the two
essential ingredients in the DLVO theory of colloidal stability \cite{VO,LesHouches}. Electrostatic interactions 
 in the presence of mobile counterions are standardly 
described by the Poisson-Boltzmann theory \cite{VO,andelman95} that embodies the {\em mean-field} approach to Coulomb
fluids and  leads to pronounced repulsive interactions between like-charged macroions. 
It has recently been realized however  
\cite{Bloom,rau,Guld84,podgornik-1,podgornik-2,podgornik-3,attard,Kjellander92,Rouzina96,Korny,Gron97,Gron98,Pincus98,AllahyarovPRL,Podgornik98,Kardar99,LinsePRL,Arenzon99,Shklovs99,Lau01,netz-SC,grosberg,Naji_epl04,ali-review}, 
that in the presence of polyvalent counterions,
electrostatic interactions can mediate strong attractive interactions 
between like-charged macroions. This attraction can not be captured by
the mean-field approach and a new paradigm dubbed the {\em strong-coupling limit} \cite{netz-SC,grosberg} was devised to
describe the equilibrium properties of Coulomb fluids when the mobile counterion charges become large. 
The transition from the mean-field  Poisson-Boltzmann description to the strong-coupling limit
 is governed by a single dimensionless electrostatic coupling parameter, being a ratio of two lengths, namely, the  Bjerrum length,
 which identifies Coulombic  interaction between ions themselves, and the Gouy-Chapman length, which describes  
electrostatic interaction between the ions 
and the charged macroion surface. This ratio involves the charge valency of  counterions, 
the dielectric constant of medium and the surface charge density of interacting macroions \cite{ali-review}. The emerging
picture of equilibrium properties of Coulomb fluids has thus become much richer than conveyed for many years by the DLVO
paradigm.

The strong-coupling attraction is not just a refinement of the mean-field
description. In fact it reverses the role of Coulomb interactions in the DLVO
theory \cite{ali-review}. Instead of stabilizing the charged macroions they act more like Lifshitz-van
der Waals interactions that tend to collapse them. Since Coulomb
interactions in the strong-coupling limit  are
much stronger than the Lifshitz-van der Waals interactions, they themselves govern the
destabilizaition of the colloids.  The collapse of a highly charged polyelectrolyte, such as DNA, in 
the presence of polyvalent counterions is the most drammatic example
of unexpected and counter-intuitive features of the strong-coupling electrostatics  \cite{Bloom,grosberg}. 
Though other physical mechanisms, such as hydration layer complementarity \cite{rau},
certainly contribute fine details to this collapse 
and account for its molecular specificity, electrostatic correlations involved in the strong-coupling limit provide
the universal background for this intriguing phenomenon. 
The elucidation of this collapse in terms of the strong-coupling electrostatic interactions represents 
one of the major recent advances in the theory of colloid stability.

In what follows, we shall concentrate on yet another facet of Coulomb
interactions in charged colloids, namely, the {\em quenched disordered} distribution of
surface charges. Usually the charge on the surfaces of the
macroions on the Poisson-Boltzmann or the strong-coupling level is assumed to be
homogeneous and constant. This is in general quite a severe assumption and
there are well known cases where this assumption is not realistic at all \cite{netz-disorder}.
Random polyelectrolytes and polyampholytes present one such case
\cite{andelman-disorder,kantor-disorder1,kantor-disorder2}. There the sequence of charges can be
distributed along the polymer backbone in a disordered manner. The Coulomb (self-)interactions 
of such polyelectrolytes are distinct and different from homogeneously charged polymers.

A case even closer to the present line of reasoning are investigations of interactions between solid surfaces in
the presence of charged surfactants. The aggregation of  surfactants at
solid surfaces in aqueous solutions was investigated with atomic force microscopy
and shows structures consistent with  half-cylinders on crystalline
hydrophobic substrates for quaternary ammonium surfactants (above the critical
micelle concentration), full cylinders on mica, and spheres on amorphous
silica \cite{manne1,manne2}. Such interfacial aggregates whose emergence and
structural details depend on the method of preparation result
from a compromise between the natural free curvature as defined by
intermolecular interactions and the constraints imposed by specific
surfactant-surface interactions and can  pattern interacting surfaces at
nanometer-length scales. Similar interfacial structures are also seen for interacting
hydrophilic mica surfaces in the presence of cetyl-trimethyl-amonium bromide
(CTAB) \cite{klein}. The surfaces appear to be covered by a
mosaic of positively and negatively charged regions and experience a strong,
{\em long-ranged attraction} which is comparable in magnitude to that between
hydrophobic surfaces, and  is orders of magnitude larger than that expected from
Lifshitz-van der Waals forces \cite{klein}.  The patterning of interacting surfaces described above is
highly disordered, depends on the method of preparation and has basic
implications also for the forces that act between other types of hydrophilic
surfaces with mixed charges, most notably the surfaces of cells and proteins, as well as in
synthetic systems. 

Motivated by these observations we set ourselves to investigate the effect of
quenched disordered surface charge, specified by a constant
mean value and mean-square deviation, on the interactions
between charged surfaces in ionic solutions. We formulate the general partition
function of a system of charged objects with a Gaussian-distributed quenched surface 
charge in a  one-component Coulomb fluid using field-theoretical techniques and the standard replica trick.  
As a particular case, we focus on the effective interaction between two apposed, ideally
polarizable planar surfaces of equal mean and mean-square deviation for disordered surface charge density. 
An explicit form for the interaction free energy can be obtained on the Poisson-Boltzmann level as
well as on the strong-coupling level. The former, interestingly enough,
shows no effect of the quenched charge disorder. The interaction free energy
at the mean-field level plus quadratic (Gaussian) fluctuations around the mean-field solution is
exactly the same as the one for a uniform charge distribution  characterized solely by a
constant mean surface charge. In the strong-coupling limit, the situation is
altogether different. Here the quenched surface charge disorder contributes
an additive {\em attractive} tail to the interaction determined by the mean
surface charge. This attractive contribution is linear in the {\em disorder coupling strength}, which
is proportional to the  mean-square deviation of
the charge distribution, and is a direct analogon 
of the electrostatic coupling parameter introduced in the case of no charge disorder \cite{ali-review}. 
This attractive tail can completely overwhelm the
entropic repulsion due to counterion confinement on the strong-coupling level
and can thus completely eliminate the finite equilibrium distance between the surfaces in this limit  \cite{netz-SC}. 
In other words, the introduction of charge disorder can destabilize the closely-packed  bound state of 
two uniformly and like-charged surfaces \cite{netz-SC,ali-review}, suggesting  a collapse transition.   
This is an important result 
because especially in the biological context the surface charge disorder is ubiquitous, stemming not only or not at
all from the method of preparation of the macroions, but from their intrinsic molecular disorder, dictated
eventually by the evolutionary processes. 

The outline of the paper is as follows: First we enunciate the results, which
is indeed quite simple, but can only be derived by fairly detalied and
rather involved technical arguments. The general formulation of the partition
function for a one-component Coulomb fluid between charged planar surfaces is
developed next in the form, where the trace over the quenched  disorder has
already been evaluated exactly on the replica level. This very complicated
partition function in the replica space can then be explicitly traced over local
electrostatic field fluctuations on the Poisson-Boltzmann level as well as on the
strong-coupling level. This can be accomplished by noting the properties of a
certain type of symmetric matrices. 
The formalism developed in the present work is quite general and  can be applied to investigate  charge disorder effects 
in a variety of systems, including interactions between spherical and  cylindrical macroions, as well.

\section{Main results}

Since the derivation of our main result contains several rather subtle and highly non-trivial
technical points  we enunciate it first stripped of all technicalities. 
Our model system is composed of two ideally polarizable surfaces located at $z = \pm a$,
\begin{figure}[t]
\centerline{\includegraphics[height=6.5cm]{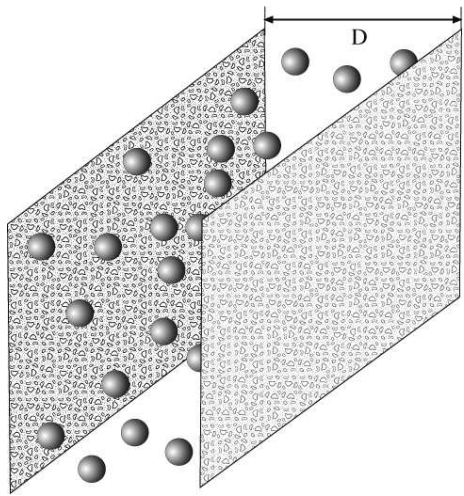}}
\caption{Schematic representation of our model system: two ideally polarizable planar surfaces with quenched surface charge 
disorder at separation $D$ with mobile (point-like) counterions of charge $Z e_0$ confined in the inter-surface space. The 
surface charge density is characterized by a Gaussian distribution with fixed mean value of magnitude $\sigma$
and mean-square deviation $g$.}
\label{fig0}
\end{figure} 
with the inter-surface spacing $D = 2a$ (see Fig. \ref{fig0}). The mean surface charge density on
both surfaces is the same with a magnitude equal to $ \sigma$, while the  mean-square deviation in surface charge density 
due to quenched disorder is given by $g$. The 
one-component Coulomb fluid between the two surfaces is composed 
of counterions of charge $Z e_{0}$, where
$Z$ is the counterion valency and $e_0$ is the elementary charge. 
For uniformly charged surfaces ($g=0$), the electrostatic coupling parameter that 
identifies the strength of electrostatic correlations in the system is given by \cite{ali-review}
\begin{equation}
 \Xi = 2\pi Z^3 \ell_{\mathrm{B}}^2 \sigma = \frac{Z^{2} \ell_{\mathrm{B}}}{\mu},
\label{xi}
\end{equation}
where 
$\ell_{\mathrm{B}}= \beta e_0^2/(4\pi \epsilon\epsilon_{0})$ is the Bjerrum length (with $\beta=1/k_{\mathrm{B}} T$) 
and $\mu= e_0/(2 \pi \ell_{\mathrm{B}} Z \sigma)$ is the Gouy-Chapman length associated with the mean charge.   
As was demonstrated by Netz \cite{netz-SC} for the {\em undisordered case}, 
the partition function of the one-component Coulomb fluid has
two different fixed points characterized by the value of $\Xi$. For
 $\Xi \ll 1$, the partition function is given by the Poisson-Boltzmann (mean-field) fixed point.
 Fluctuations  around this fixed point may be accounted for by a systematic 
 loop expansion \cite{podgornik-1,podgornik-2,attard,netz-SC}. 
At elevated coupling,  the loop expansion scheme breaks down  and the  
partition function for  $\Xi \gg 1$  is determined  by the asymptotic strong-coupling theory, which
corresponds to the leading term of a virial expansion, and is in fact 
given  by single-particle contributions.

In the presence of quenched disorder ($g>0$), 
one can follow the same line of reasoning, 
by introducing the mean electrostatic coupling parameter $\Xi$ in analogy
with Eq. (\ref{xi}) but with $\sigma$ now being the {\em mean} surface charge density.
In the mean-field limit, characterized by $\Xi \ll 1$, we demonstrate  that  the disorder makes {\em no}
contribution to the interaction free energy, if the bounding surfaces are
ideally polarizable and thus do not allow the penetration of the field. The 
total interaction pressure in this limit is given by the
standard mean-field (Poisson-Boltzmann) expression $P_{\mathrm{MF}}(D)$ and an additive
contribution, $P_{2}(D)$,  due to quadratic (Gaussian) fluctuations around the  mean-field solution  as
\begin{equation}
P(D) = P_{\mathrm{MF}}(D) + P_{2}(D).
\end{equation}
Both terms depend only on the mean surface charge density $\sigma$ and have been thoroughly
analyzed before \cite{VO,podgornik-1,attard,Pincus98,Kardar99,netz-SC}.
The Poisson-Boltzmann limit  for ideally polarizable bounding interfaces is thus
unaffected by the presence of disorder, {\em i.e.} it does not depend on $g$. This result is completely consistent with 
conclusions reached in Ref. \cite{netz-disorder} {\em via} an altogether  different route.

In the strong-coupling limit, characterized by large mean electrostatic
coupling parameter $\Xi \gg 1 $,  however, the interaction free energy is obtained to contain an
essential contribution stemming from the quenched surface charge disorder. The
total interaction pressure is given by the sum of the standard  strong-coupling pressure
$P_{\mathrm{SC}}(D)$ \cite{netz-SC} that depends only on the mean charge density $\sigma$ plus a
disorder contribution, $P_\chi(D)$,  depending on $g$, i.e. 
\begin{equation}
P(D) = P_{\mathrm{SC}}(D) +P_\chi(D), 
\label{dis-1}
\end{equation}
where  the disorder term in dimensionless representation reads 
\begin{equation}
   \frac{\beta P_\chi(D)}{2\pi \ell_{\mathrm{B}}\sigma^2} = - {\chi}\bigg(\frac{2\mu}{D}\bigg), 
\end{equation}
which is attractive and linear in $\chi$  defined as 
\begin{equation}
\chi = \frac{Z^{2} (\beta e_{0})^{2} g}{8 \pi (\epsilon\epsilon_{0})^{2}} = 2\pi Z^2\ell_{\mathrm{B}}^2 g.
\label{chi}
\end{equation}
Clearly by comparing Eq. (\ref{xi}) and Eq. (\ref{chi}), we can claim that $\chi$
represents the dimensionless {\em disorder coupling parameter}.
$P_{\mathrm{SC}}(D)$  has been analyzed thoroughly before \cite{netz-SC}. The disorder term in Eq. (\ref{dis-1})
incidentaly has the same scaling form as the one-particle ideal entropy term in
$P_{\mathrm{SC}}(D)$. It can change the sign of the interactions at small values of the
inter-surface spacing $D$, as will be shown later. 
The quenched charge disorder thus affects only the strong-coupling limit in a rather simple way,
contributing an additive term to the free energy or the interaction pressure.
After this introductory exposition, we next derive these results in full glory and all the relevant technical
details.

\section{General formalism: The replica method for quenched averaging}

The partition function of a one-component (counterion-only) Coulomb fluid in the
field of an external fixed charge charaterized by the volume charge density $\rho(\vct
r)$ can be derived in the form of a functional integral over the fluctuating
electrostatic field $\phi(\vct r)$ as \cite{podgornik-1, podgornik-2}
\begin{equation}
{\mathcal Z} = e^{- {\ul12} \ln\det{\beta v(\vct r, \vct r')}}~\int {\cal
D}[\phi(\vct r)] ~e^{- \beta{\cal H}[\phi(\vct r)]},
\end{equation}
where the field-action is given by
\begin{equation}
\beta{\cal H}[\phi(\vct r)] = {\ul12} \beta \int\!\!\int\!\! {\mathrm{d}} \vct
r {\mathrm{d}} \vct r'~ \phi(\vct r) v^{-1}(\vct r, \vct r') \phi(\vct r')  - 
\tilde{\lambda} \int\!\! {\mathrm{d}} \vct r ~\Omega (\vct r)~e^{- i \beta Z e_{0} \phi(\vct r)} + i
\beta \int\!\! {\mathrm{d}}\vct r~\rho(\vct r) \phi(\vct r). 
\label{act-1}
\end{equation}
Here the geometry function $\Omega (\vct r)$ specifies the volume accessible for counterions
in space (it, for instance, takes into account the presence of hard walls). 
The rescaled absolute activity (fugacity), $\tilde \lambda$ (connected to the chemical potential, $\mu$, 
{\em via} $\mu = \ln{\lambda}$), is defined as
\begin{equation}
\tilde{\lambda} = \lambda ~e^{{\ul12} e^{2}_{0} \beta v(\vct r, \vct r)}.
\label{lambda-tilde-def}
\end{equation}
The Coulomb interaction potential $v(\vct r, \vct r') = 1/(4\pi \epsilon \epsilon_{0} \vert \vct r -
\vct r' \vert)$ is a solution of
\begin{equation}
-\epsilon \epsilon_{0}
\nabla^{2} v(\vct r, \vct r') = \delta(\vct r - \vct r').
\end{equation}
and thus obviously  the inverse Coulomb operator is given by
\begin{equation}
v^{-1}(\vct r, \vct r') =  -\epsilon \epsilon_{0}
\nabla^{2}\delta(\vct r - \vct r').
\end{equation}
We assume in what follows that the external charge distribution
has a quenched disordered component and is in fact distributed with a Gaussian
probability distribution around its mean $\rho_{0}(\vct r)$ as
\begin{equation}
{\mathrm{const.}} \times e^{- \ul12 \int
{\mathrm{d}}\vct r~g^{-1}(\vct r) \left( \rho(\vct r) - \rho_{0}(\vct r)\right)^{2}}
\label{distr-1}
\end{equation}
Here the width of the charge
disorder distribution, {\em i.e.} the mean-square charge density is given by $g(\vct r) $.
The average over quenched charge disorder is obtained by applying the standard
Edwards-Anderson replica {\em ansatz} \cite{orland} in the form
\begin{equation}
{\cal F} = - k_{\mathrm{B}} T ~\overline{\log{{\mathcal Z}}} = - k_{\mathrm{B}} T~\lim_{n \rightarrow 0}
\frac{\overline{{\mathcal Z}^{n}} - 1}{n}, 
\label{free-1}
\end{equation}
where the disorder average is defined with respect to the external charge
density distribution, Eq. (\ref{distr-1}), as
\begin{equation}
\overline{(\dots)} = \int {\cal D}[\rho(\vct r)] (\dots) e^{- \ul12 \int
{\mathrm{d}}\vct r~g^{-1}(\vct r) \left( \rho(\vct r) - \rho_{0}(\vct r)\right)^{2}}.
\end{equation}
The quenched average over the charge density disorder affects only the source term in 
Eq. (\ref{act-1}). Thus we only need to evaluate 
\begin{eqnarray}
\overline{e^{- i \beta \int {\mathrm{d}}\vct r~ \rho(\vct r) \phi(\vct r)}} &=& 
\int {\cal D}[\rho(\vct r)] ~e^{\textstyle - i \beta \sum_{\alpha} \int {\mathrm{d}}\vct
r~ \rho(\vct r)\phi_{\alpha}(\vct r) - \ul12 \int {\mathrm{d}}\vct r~g^{-1}(\vct r)
\left( \rho(\vct r) - \rho_{0}\right)^{2}} = \nonumber\\
&=& {\mathrm{const.}} \times e^{- \ul12 \beta^{2} \int {\mathrm{d}}\vct r~g(\vct r) \sum_{\alpha,
\beta} \phi_{\alpha}(\vct r) \phi_{\beta}(\vct r) - i  \beta \sum_{\alpha}
\int {\mathrm{d}}\vct r~ \rho_{0}(\vct r)\sum_{\alpha} \phi_{\alpha}(\vct
r)}\nonumber\\
~
\end{eqnarray}
for $\alpha, \beta=1,\ldots, n$ being the replica labels. 
Taking this into account, the final form of the replicated 
partition function $\overline{{\mathcal Z}^{n}}$ can be obtained as
\begin{eqnarray}
\overline{{\mathcal Z}^{n}} = e^{- {\ul12} n \ln\det{\beta v(\vct r, \vct
r')}}~\int {\cal D}[\phi_{\alpha}(\vct r)] ~e^{- \beta \tilde{\cal H}[\phi_{\alpha}(\vct
r)]},
\label{rep-1}
\end{eqnarray}
with
\begin{equation}
\beta \tilde{\cal H}[\phi_{\alpha}(\vct r)] = {\ul12} \sum_{\alpha \beta}\!\!
\int\!\!\!\int {\mathrm{d}}\vct r {\mathrm{d}}\vct  r' ~{\cal D}_{\alpha
\beta}(\vct r, \vct r')\phi_{\alpha}(\vct
r)\phi_{\beta}(\vct r) - V[\phi_{\alpha}(\vct r)]  + i 
\beta \int {\mathrm{d}}\vct r ~\rho_{0}(\vct r)\sum_{\alpha} \phi_{\alpha}(\vct r),
\label{H-1}
\end{equation}
where 
\begin{equation}
V[\phi_{\alpha}(\vct r)] = \tilde{\lambda} \int\!\! {\mathrm{d}} \vct r ~\Omega (\vct r)~\sum_{\alpha} e^{-
i \beta Z e_{0} \phi_{\alpha}(\vct r)}.
\end{equation}
Above we introduced the following matrix in the replica space
\begin{equation}
{\cal D}_{\alpha \beta}(\vct r, \vct r') = \left( \beta v^{-1}(\vct r, \vct r')
 +\beta^{2} g(\vct r) \delta(\vct r - \vct r') \right)\delta_{\alpha
\beta} + \beta^{2} g(\vct r) \delta(\vct r - \vct r')  (1 - \delta_{\alpha
\beta}).
\label{def-1}
\end{equation}
The  expression (\ref{rep-1}) together with Eq. (\ref{free-1}) represents the
starting formulation for the free energy in the presence of quenched charge
disorder. There are two separate problems with the above replicated partition
function  (\ref{rep-1}) that we have to resolve. First of all there is the non-linear
term in the field-action that precludes direct integration. On top of that we
have to evaluate the functional integral in the replica space for arbitrary number of replicas, $n$, 
and then take the limit of $n \rightarrow 0$. 

The way we will approach this rather formidable task is to combine the methods
developed for the one component (counterion-only) Coulomb fluid without disorder
\cite{netz-SC} and modify them as we proceed to incorporate appropriately the
disorder effects. We shall start with the Poisson-Boltzmann or the saddle-point
limit, which is exact for $\Xi\rightarrow 0$ \cite{netz-SC}, and then proceed to the strong-coupling limit.

\section{The mean-field  (Poisson-Boltzmann)  limit}

Let us first investigate the saddle-point limit of the functional integral (\ref{rep-1}). 
The saddle-point equation for the $\phi_{\alpha}(\vct r)$ field is given
obviously by
\begin{equation}
\int {\mathrm{d}} \vct r' \sum_{\beta}{\cal D}_{\alpha \beta}(\vct r, \vct r')  \phi_{\beta}(\vct
r') - \frac{\partial V[\phi_{\alpha}(\vct r)]}{\partial \phi_{\alpha}(\vct r)}
+ i \beta \rho_{0}(\vct r) = 0.
\end{equation}
Taking into account Eq. (\ref{def-1}), this amounts to
\begin{equation}
-\beta \epsilon\epsilon_{0} \nabla^{2} \phi_{\alpha}(\vct r) + \beta^{2}
g(\vct r) \sum_{\beta} \phi_{\beta}(\vct r) - \frac{\partial V[\phi_{\alpha}(\vct r)]}{\partial \phi_{\alpha}(\vct r)}
+ i \beta \rho_{0}(\vct r)= 0.
\end{equation}
In the replica formulation we have to take the limit $n \rightarrow 0$,
which implies that 
\begin{equation}
\lim_{n \rightarrow 0} \sum_{\beta} \phi_{\beta}(\vct r) \rightarrow
0. 
\end{equation}
By furthermore making the substitution $\phi_{\alpha}(\vct r) \rightarrow i
\phi_{\alpha}(\vct r)$ and noting that in the limit $n \rightarrow 0$, 
the index $\alpha$ becomes irrelevant, and denoting the corresponding
potential as $\phi_{\mathrm{MF}}(\vct r)$, we are left with
\begin{equation}
\epsilon\epsilon_{0} \nabla^{2} \phi_{\mathrm{MF}}(\vct r)  + \tilde \lambda 
Z e_{0} ~\Omega (\vct r)~e^{- \beta Z e_{0} \phi_{\mathrm{MF}}(\vct r)} = -\rho_{0}(\vct r).
\label{PB-1}
\end{equation}
This is of course nothing but the Poisson-Boltzmann equation for the fixed mean 
charge density $\rho_{0}(\vct r)$ and the mean-field Poisson-Boltzmann potential 
$\phi_{\mathrm{MF}}(\vct r)$ \cite{podgornik-1, netz-SC}. The corresponding mean-field
(saddle-point) action can be written as
\begin{equation}
\beta  \tilde{\cal H}[\phi_{\mathrm{MF}}(\vct r)] = - {\ul12} \epsilon\epsilon_0 \beta\int {\mathrm{d}} \vct r ~ (\bnabla 
 \phi_{\mathrm{MF}}(\vct r))^2 - \tilde{\lambda} \int {\mathrm{d}} \vct r ~ \Omega (\vct r)~e^{-
\beta Z e_{0} \phi_{\mathrm{MF}}(\vct r)} + 
\beta \int {\mathrm{d}}\vct r~\rho_0(\vct r) \phi_{\mathrm{MF}}(\vct r).
\end{equation}
The effect of the quenched charge disorder on
the mean  potential is thus nil. Obviously the saddle-point action can be
taken out of the functional integral since it is simply multiplicative in the
number of replicas.

We now discuss the fluctuations around the mean-field, or the
saddle-point,  fields. Expanding $\beta \tilde{\cal H}[\phi_{\alpha}(\vct r)]$ in 
Eq. (\ref{H-1}) to the second order in the deviations from $\phi_{\mathrm{MF}}(\vct r)$, 
we remain with
\begin{eqnarray}
\overline{{\mathcal Z}^{n}} = e^{- {\ul12} n \ln\det{\beta v(\vct r, \vct
r')} - n \beta \tilde{\cal H}_{\mathrm{MF}}[\phi_{\mathrm{MF}}(\vct r)]}~\int {\cal
D}[\phi_{\alpha}(\vct r)] ~e^{- \beta \tilde{\cal H}_{2}[\phi_{\alpha}(\vct r)]},
\label{rep-2}
\end{eqnarray}
where $ \tilde{\cal H}_{\mathrm{MF}}[\phi_{\mathrm{MF}}(\vct r)] =  \tilde{\cal H}[\phi_{\mathrm{MF}}(\vct r)]$, 
\begin{equation}
\beta \tilde{\cal H}_{2}[\phi_{\alpha}(\vct r)] = {\ul12} \sum_{\alpha \beta}
\int\!\!\int {\mathrm{d}}\vct r {\mathrm{d}}\vct r' \phi_{\alpha}(\vct r) {\cal  G}_{\alpha
\beta}(\vct r, \vct r') \phi_{\beta}(\vct r),
\end{equation}
and 
\begin{equation}
{\cal  G}_{\alpha \beta}(\vct r, \vct r') = {\cal D}_{\alpha \beta}(\vct r, \vct r')
- \frac{\partial^{2} V[\phi_{\alpha}(\vct r)]}{\partial \phi^{2}_{\alpha}(\vct
r)}\bigg|_{\phi_{\mathrm{MF}}} \delta_{\alpha \beta} \delta(\vct r - \vct r').
\end{equation}
In the above expression the second term has to be evaluated at the saddle-point
and thus yields
\begin{equation}
\frac{\partial^{2} V[\phi_{\alpha}(\vct r)]}{\partial \phi^{2}_{\alpha}(\vct
r)}\bigg|_{\phi_{\mathrm{MF}}}  = - \tilde\lambda (\beta Z e_{0})^{2} ~\Omega (\vct r)~e^{- \beta Z e_{0} \phi_{\mathrm{MF}}(\vct r)}.
\end{equation}
Thus we obtain explicitly
\begin{equation}
{\cal  G}_{\alpha \beta}(\vct r, \vct r') = \left[ \beta v^{-1}(\vct r, \vct r')
 + \left(  \tilde\lambda (\beta Z e_{0})^{2} ~\Omega (\vct r)~e^{- \beta Z e_{0} \phi_{\mathrm{MF}}(\vct r)}
 + \beta^{2} g(\vct r) \right)\delta(\vct r - \vct r')\right]  \delta_{\alpha
\beta} + \beta^{2}
g(\vct r) \delta(\vct r - \vct r')  (1 - \delta_{\alpha \beta}).
\end{equation}
Since $\beta \tilde{\cal H}_{2}[\phi_{\alpha}(\vct r)] $ is Gaussian in
$\phi_{\alpha}(\vct r)$ the functional integral,  Eq. (\ref{rep-2}),  can be
evaluated explicitly yielding 
\begin{eqnarray}
\overline{{\mathcal Z}^{n}} = e^{- \ul12 n \ln\det{\beta v(\vct r, \vct
r')} - n \beta \tilde{\cal H}_{\mathrm{MF}}[\phi_{\mathrm{MF}}(\vct r)] - {\ul12}\ln\det{{\cal
G}_{\alpha\beta}(\vct r, \vct r')}}.
\label{rep-2b}
\end{eqnarray}
The task now is to evaluate the determinant $\det{{\cal G}_{\alpha \beta}(\vct r,
\vct r')}$ explicitly as a function of $n$ and then use this to evaluate the
$n \rightarrow 0$ limit. This can be done
with the help of the following matrix identity. Take  a symmetric $n \times n$ matrix of the
general form
\begin{equation}
M_{\alpha \beta} = 
\left[ 
\begin{array}{llll} 
b  & s  & s  &\dots  \\
s  & b  & s  &\dots  \\
s & s  & b  &\dots   \\
\dots & \dots  & \dots  & \dots \\
\end{array}\right] = b ~\delta_{\alpha \beta} + s ~( 1 - \delta_{\alpha
\beta}).
\end{equation}
It is rather straightforward to show {\em via} induction that
\begin{equation}
\det{M_{\alpha \beta}} = (b-s)^{n}\left(1 + \frac{n s}{b-s}\right).
\end{equation}
In order to exploit this matrix identity we first define the operator $G(\vct
r, \vct r')$ as
\begin{equation}
\int {\mathrm{d}} \vct r'' \left( \beta v^{-1}(\vct r, \vct r'')
 + \tilde\lambda (\beta Z e_{0})^{2} ~\Omega (\vct r)~e^{- \beta Z e_{0} \phi_{\mathrm{MF}}(\vct r)}
\delta(\vct r - \vct r'')\right) G(\vct
r'', \vct r') = \delta(\vct r - \vct r'),
\end{equation}
or equivalently 
\begin{equation}
\left( -\beta \epsilon\epsilon_{0} \nabla^{2}  + \tilde\lambda (\beta Z e_{0})^{2}
~\Omega (\vct r)~e^{- \beta Z e_{0} \phi_{\mathrm{MF}}(\vct r)}\right) G(\vct
r, \vct r') = \delta(\vct r - \vct r').
\end{equation}
With this definition and by using the ${\rm Tr} \log$ formula, the determinant 
$\det{{\cal G}_{\alpha \beta}(\vct r, \vct r')}$ can be obtained explicitly as
\begin{equation}
{\ul12}\ln\det{{\cal G}_{\alpha \beta}(\vct r, \vct r')} = - {\textstyle\frac{n}{2}}
\ln\det{{G}(\vct r, \vct r')} + {\ul12} {\rm Tr}\ln{\left(\delta(\vct r - \vct r') + n \beta^{2} 
g(\vct r) G(\vct r, \vct r')\right)}.
\end{equation}
The final expression for the replicated partition function becomes 
\begin{eqnarray}
\ln{\overline{{\mathcal Z}^{n}}} = - {\textstyle\frac{n}{2}} \ln\det{\!\int\!\!
{\mathrm{d}}\vct r'\beta v(\vct r, \vct r') {G}^{-1}(\vct r',
\vct r'')} - n \beta \tilde{\cal H}_{\mathrm{MF}}[\phi_{\mathrm{MF}}(\vct r)]  - {\ul12} {\rm Tr}
\ln{\left(\delta(\vct r - \vct r') + n \beta^{2}  g(\vct r) G(\vct r, \vct
r')\right)}.
\label{rep-2c}
\end{eqnarray}
We have thus evaluated the partition function on the Poisson-Boltzmann level plus
Gaussian-field fluctuations {\em exactly} as a function of $n$. The
replica trick now leads directly to the free energy, Eq. (\ref{free-1}), of the
form
\begin{eqnarray}
{\cal F}_\lambda &=& - k_{\mathrm{B}} T~\lim_{n \rightarrow 0}
\frac{\overline{{\mathcal Z}^{n}} - 1}{n} = \nonumber\\
&=& \tilde{\cal H}_{\mathrm{MF}}[\phi_{\mathrm{MF}}(\vct r)] + {\textstyle\frac{k_{\mathrm{B}} T}{2}} \ln\det{\!\int\!\!
{\mathrm{d}}\vct r'\beta v(\vct r, \vct r') {G}^{-1}(\vct r', \vct r'')} +
{\textstyle\frac{\beta}{2}} {\rm Tr}~ g(\vct r) G(\vct r, \vct r').
\label{interm-fin-1}
\end{eqnarray}
These three terms have a straightforward physical interpretation
\cite{podgornik-3, netz-SC}. The first term is the standard mean-field result, the second
is the one-loop correction  around the mean-field, and the third one is the contribution  of
the disorder. Obviously the latter is linear in the strength of the disorder $g$.

We now go back from the grand-canonical to the canonical partition function {\em via}
the Legendre transform \cite{netz-SC}
\begin{equation}
{\cal F}_N = {\cal F}_\lambda + k_{\mathrm{B}} T~N~\ln{\lambda}
\label{legendre-F_N}
\end{equation}
where $N$ is the total number of counterions, which can be expressed through the absolute activity, $\lambda$, as
\begin{equation}
N = -\lambda \frac{\partial {\cal F}_\lambda}{\partial \lambda} = - \tilde \lambda \frac{\partial {\cal
F}(\tilde \lambda)}{
\partial \tilde \lambda}.
\label{legendre-N-lambda}
\end{equation}
(This simply means that the $\log$ of the absolute activity is defined up to an additive constant.) To the lowest order, and on the saddle-point level, we obtain straightforwardly
\begin{equation}
N =\tilde  \lambda \int {\mathrm{d}}\vct r ~\Omega (\vct r)~e^{- \beta Z e_{0} \phi_{\mathrm{MF}}(\vct r)}. 
\end{equation}
This relation is nothing but the normalization condition for the mean-field density profile of counterions, 
$\tilde \lambda~\Omega (\vct r)~e^{- \beta Z e_{0} \phi_{\mathrm{MF}}(\vct r)}$,  that can be connected with the electroneutrality condition {\em via} $2 \sigma S =  N Z e_{0}$, where $S$ is the total area and $\sigma$ the surface charge density of macroions. Thus we finally remain with
\begin{equation}
{\cal F}_N = {\cal F}_\lambda - k_{\mathrm{B}} T~ N \ln{\int {\mathrm{d}}\vct r
~\Omega (\vct r)~e^{- \beta Z e_{0} \phi_{\mathrm{MF}}(\vct r)}} + k_{\mathrm{B}} T~ N \ln{N}.
\label{interm-fin}
\end{equation}
For two charged planar surfaces, for instance, 
the interaction pressure can be obtained by differentiating  the free energy ${\cal F}_N$ 
with respect to the separation, $D$.

\section{Mean-field results for two charged plates}

Let us assume now that our system is composed of two planar surfaces, located at
 $z = \pm a$,  with mean surface charge density $-\sigma$. It is thus obviously homogeneous in directions $\brho = (x,y)$. The mean volume charge distribution therefore reads
\begin{equation}
\rho_{0}(\vct r) = \rho_{0}(z, \brho)= -\sigma \delta(z - a) -\sigma \delta(z + a).   
\end{equation}
The mean-square charge density due to disorder is analogously written as 
\begin{equation}
g(\vct r) = g(z, \brho) = g \delta(z - a) + g \delta(z + a).
\label{ans-1}
\end{equation}
Note also that since counterions are confined in the inter-surface space (Fig. \ref{fig0}), 
the geometry function $\Omega (\vct r)$ reads
\begin{equation}
   \Omega (\vct r) =   \Omega (z)  =\left\{
   \begin{array}{ll}
    1  & {\,\,\,\,\,-a<z<a} \\
    \\
    0   & {\,\,\,\,\,\mathrm{otherwise}}.
   \end{array}
   \right.
   \label{eq:omega}
\end{equation}
From Eqs. (\ref{interm-fin-1}) and (\ref{interm-fin}) we derive the corresponding canonical  free energy in the form
\begin{equation}
{\cal F}_N =  {\cal F}_N(g = 0) + {\textstyle\frac{\beta}{2}} g \!\!\int\!\! {\mathrm{d}}^{2}\brho
\left(  G(a, a; \brho, \brho) + G(-a, -a; \brho, \brho) \right),
\label{res-11}
\end{equation}
where we decoupled the disorder independent terms, ${\cal F}_N(g = 0)$, from those that depend on $g$. The former contribution 
may be written as ${\cal F}_N(g = 0)=  {\cal F}_N^{\mathrm{MF}}[ \phi_{\mathrm{MF}}(\vct r)] +  {\cal F}_N^{(2)} $, where 
\begin{equation}
   {\cal F}_N^{\mathrm{MF}} [ \phi_{\mathrm{MF}}(\vct r)]\!= \!\!\int\! {\mathrm{d}} \vct r~ \bigg(\!-\frac{\epsilon\epsilon_0}{2} (\bnabla 
 \phi_{\mathrm{MF}}(\vct r))^2 + 
  \rho_0(\vct r) \phi_{\mathrm{MF}}(\vct r) \bigg) - k_{\mathrm{B}} T~ N \ln{\int \!{\mathrm{d}}\vct r
~\Omega (\vct r)~e^{- \beta Z e_{0} \phi_{\mathrm{MF}}(\vct r)}} + k_{\mathrm{B}} T~ (N \ln{N}-N).
\label{MF-free}
\end{equation}
is the mean-field Poisson-Boltzmann free energy, and $ {\cal F}_N^{(2)}  = {\textstyle\frac{k_{\mathrm{B}} T}{2}} \ln\det{\!\int\!
{\mathrm{d}}\vct r'\beta v(\vct r, \vct r') {G}^{-1}(\vct r', \vct r'')}  $ is the contribution of Gaussian fluctuations around the 
mean-field solution. These terms have already been
evaluated elsewhere \cite{podgornik-1} and will not be analyzed here. 

If we assume furthermore that the mean electrostatic field is confined solely to
the region between the surfaces, that is assuming that the surfaces  are ideally polarizable, then
the last term in Eq. (\ref{res-11}) describing the effects of the disorder is identically zero since it is
straightforward to show that in this case $$
G(a, a; \brho, \brho) = G(-a, -a; \brho, \brho) = 0.$$
Therefore, with ideally polarizable bounding surfaces, the
effect of the disorder on the electrostatic interactions is zero also 
on the one-loop level.
Evaluating the interaction pressure between the two apposed surfaces from 
Eq. (\ref{res-11}), we thus find
\begin{equation}
P(D) = P_{\mathrm{MF}}(D) + P_{2}(D).
\end{equation}
The first component of pressure is obtained from the mean-field Poisson-Boltzmann free energy (\ref{MF-free}), 
which is always repulsive.  For two planar equally charged surfaces at large separations and in dimensionless units, 
one has \cite{VO,podgornik-1}
\begin{equation}
 \tilde P(D)=   \frac{\beta P_{\mathrm{MF}}(D)}{2\pi \ell_{\mathrm{B}}\sigma^2} \simeq \bigg(\frac{\pi\mu}{D}\bigg)^2. 
\end{equation}
The second  term, stemming from fluctuations, in fact corresponds to zero-frequency
(static) Lifshitz-van der Waals interaction \cite{podgornik-3}. For two plates, this term
contributes an attractive component that scales as \cite{podgornik-1,podgornik-2,attard,Pincus98,Kardar99}
\begin{equation}
 \tilde P_2(D)=   \frac{\beta P_2(D)}{2\pi \ell_{\mathrm{B}}\sigma^2} \simeq -\pi^2\bigg(\frac{\mu}{D}\bigg)^3\ln (D/\mu). 
\end{equation}

The preceding result is pleasingly consistent
with a previous study of quenched disorder on the weak-coupling level  \cite{netz-disorder}, 
which also shows no disorder contribution to the density profile of counterions at a single charged surface.

\section{The strong-coupling limit}

Now we address the strong-coupling limit, $\Xi\gg 1$, which was shown to correspond to the
first order term in the expansion of the partition function in terms of the
absolute activity \cite{netz-SC}. Expanding Eq. (\ref{rep-1}) in terms of
$\tilde\lambda$ we obtain to the lowest order
\begin{eqnarray}
\overline{{\mathcal Z}^{n}} &=&  \overline{{\mathcal Z}^{n}_{0}} + \tilde\lambda~
\overline{{\mathcal Z}^{n}_{1}} + {\cal O}(\tilde\lambda^{2}) = \nonumber\\
&=& e^{- {\ul12} n \ln\det{\beta v(\vct r,
\vct r')}}~\left( \int {\cal D}[\phi_{\alpha}(\vct r)] ~e^{- \beta \tilde{\cal
H}_{0}[\phi_{\alpha}(\vct r)]} + \tilde\lambda \int\!\!{\mathrm{d}}\vct R~\Omega (\vct R)\sum_{\gamma=1}^n \!\int {\cal
D}[\phi_{\alpha}(\vct r)] ~e^{-\beta \tilde{\cal H}_{1}[\phi_{\alpha}(\vct r), \phi_{\gamma}(\vct
R)]}\right),\nonumber\\
\label{SC-1}
\end{eqnarray}
with
\begin{equation}
\beta \tilde{\cal H}_{0}[\phi_{\alpha}(\vct r)] = {\ul12} \sum_{\alpha \beta}\!\!
\int\!\!\!\int {\mathrm{d}}\vct r {\mathrm{d}}\vct  r' ~{\cal D}_{\alpha
\beta}(\vct r, \vct r')\phi_{\alpha}(\vct
r)\phi_{\beta}(\vct r)  + i 
\beta \int {\mathrm{d}}\vct r ~\rho_{0}(\vct r)\sum_{\alpha} \phi_{\alpha}(\vct r),
\label{H-2}
\end{equation}
and 
\begin{equation}
\beta \tilde{\cal H}_{1}[\phi_{\alpha}(\vct r)] = {\ul12} \sum_{\alpha \beta}\!\!
\int\!\!\!\int {\mathrm{d}}\vct r {\mathrm{d}}\vct  r' ~{\cal D}_{\alpha
\beta}(\vct r, \vct r')\phi_{\alpha}(\vct
r)\phi_{\beta}(\vct r)  - i 
\beta \int {\mathrm{d}}\vct r ~\rho_{0}(\vct r)\sum_{\alpha} \phi_{\alpha}(\vct r) +
i \beta Z e_{0} \phi_{\gamma}(\vct R).
\label{H-3}
\end{equation}
${\cal D}_{\alpha \beta}(\vct r, \vct r')$ was defined already in Eq. (\ref{def-1}).
The first functional integral in Eq. (\ref{SC-1}) can be evaluated
straightforwardly by noting that it is equivalent to Eq. (\ref{rep-1}) for
$V[\phi_{\alpha}] = 0$. Since in this limit obviously $$\beta G(\vct r, \vct r') =
v(\vct r, \vct r')$$ we obtain immediately
\begin{eqnarray}
{\cal F}_{0} &=& - k_{\mathrm{B}} T~\lim_{n \rightarrow 0}
\frac{\overline{{\mathcal Z}^{n}_{0}} - 1}{n} = \nonumber\\
&=& {\ul12} \int\!\!\int {\mathrm{d}}\vct r {\mathrm{d}} \vct r' \rho_{0}(\vct r) v(\vct r,
\vct r') \rho_{0}(\vct r') +
{\ul12} {\rm Tr}~ g(\vct r) v(\vct r, \vct r')
\end{eqnarray}
corresponding to the zeroth-order term of the virial expansion. 
The first term is the standard electrostatic interaction energy between
external charges and the second term is the contribution of the disorder. We
note here again that with the {\em ansatz } (\ref{ans-1}), the disorder term does
not depend on the separation between the bounding surfaces because $v(z =a, 
z' = a: \brho, \brho)$ does not depend on $a$. 

The second term in Eq. (\ref{SC-1}) is a bit trickier but can be again straightforwardly
reduced to
\begin{equation}
\overline{{\mathcal Z}^{n}_{1}} = \overline{{\mathcal Z}^{n}_{0}} ~
\int {\mathrm{d}} \vct R~\Omega (\vct R)\sum_{\gamma=1}^n e^{\beta u_{\gamma}(\vct R) },
\label{foo-1a}
\end{equation}
where
\begin{equation}
\beta u_{\gamma}(\vct R) = \beta^{2} Z e_{0} \int {\mathrm{d}} \vct r'
\sum_{\alpha} {\cal D}^{-1}_{\alpha
\gamma}(\vct r', \vct R) \rho_{0}(\vct r') - {\ul12} (\beta Z e_{0})^{2}  {\cal
D}^{-1}_{\gamma \gamma}(\vct R, \vct R). 
\label{foo-1}
\end{equation}
The task now is to evaluate the expressions $\sum_{\alpha} {\cal D}^{-1}_{\alpha
\beta}(\vct r', \vct R)$ and ${\cal D}^{-1}_{\alpha
\alpha}(\vct R, \vct R)$ explicitly as a function of $n$. This can be done
just as before by considering the following  identity valid for a symmetric $n \times n$
matrix of the general form
\begin{equation}
M_{\alpha \beta} =  b ~\delta_{\alpha \beta} + s ~( 1 - \delta_{\alpha \beta}).
\end{equation}
One can derive {\em via} induction that
\begin{equation}
M^{-1}_{\alpha \alpha} = \frac{1}{(b-s)}\left(1
+ \frac{s}{(b-s)}  (
1 + \frac{n s}{b - s})^{-1}\right)
\end{equation}
and
\begin{equation}
\sum_{\beta} M^{-1}_{\alpha \beta} = \frac{1}{(b-s)}\left(1 + \frac{n s}{b-s}\right)^{-1}.
\end{equation}
With the help of these matrix identities we can derive that
\begin{equation}
{\cal D}^{-1}_{\gamma \gamma}(\vct R, \vct R) = \int {\mathrm{d}}\vct r ~ G(\vct R,
\vct r) T(\vct r, \vct R),
\end{equation}
where
\begin{equation}
 T(\vct r, \vct r') = \delta(\vct r - \vct r') - \beta^{2} \int {\mathrm{d}}\vct
r'' g(\vct r'') G(\vct r'', \vct r) \left( \delta(\vct r - \vct r') + n
\beta^{2} g(\vct r) G(\vct r,\vct r')\right)^{-1}.
\end{equation}
Similarly
\begin{equation}
\sum_{\alpha} {\cal D}^{-1}_{\alpha
\gamma}(\vct r', \vct R) = \int {\mathrm{d}} \vct r' G(\vct r, \vct r' )\left( \delta(\vct r - \vct r') + n
\beta^{2} g(\vct r) G(\vct r,\vct r')\right)^{-1}.
\end{equation}
Neither of the above two equalities depends on the index $\gamma$,  and the
$\gamma$-terms in Eq. (\ref{foo-1a}) are thus additive leading to 
\begin{equation}
\overline{{\mathcal Z}^{n}_{1}} = n~\overline{{\mathcal Z}^{n}_{0}}~\int {\mathrm{d}} \vct R ~\Omega (\vct R) 
~ e^{\beta u_{\gamma}(\vct R) }.
\end{equation}
Since this is already first order in $n$, we only need Eq. (\ref{foo-1}) up to the zeroth
order in $n$, thus obtaining
\begin{equation}
\overline{{\mathcal Z}^{n}} = \overline{{\mathcal Z}^{n}_{0}} \left(1 + \tilde\lambda
~n \int {\mathrm{d}} \vct R~\Omega (\vct R)~  e^{\beta u(\vct R) } \right),
\end{equation}
where
\begin{equation}
  \beta u(\vct R)  =  \beta Z e_{0} \int {\mathrm{d}} \vct r'
v(\vct r', \vct R) \rho_{0}(\vct r') + {\ul12} (\beta Z e_{0})^{2}\int {\mathrm{d}}
\vct r g(\vct r) v^{2}(\vct r, \vct R), 
\label{u_final}
\end{equation}
and hence {\em in extenso}
\begin{equation}
\overline{{\mathcal Z}^{n}} = \overline{{\mathcal Z}^{n}_{0}} \left(1 + \lambda 
~n \int {\mathrm{d}} \vct R~\Omega (\vct R) ~ e^{ \beta Z e_{0} \int {\mathrm{d}} \vct r'
v(\vct r', \vct R) \rho_{0}(\vct r') + {\ul12} (\beta Z e_{0})^{2}\int {\mathrm{d}}
\vct r g(\vct r) v^{2}(\vct r, \vct R)} \right).
\end{equation}
Note that in this formula we have the original fugacity $\lambda$ and not the rescaled value $\tilde\lambda$
(see Eq. (\ref{lambda-tilde-def})). The
divergent self-interaction is cancelled by an equal and opposite term in
$\beta u(\vct R)$.  The total free energy can now be obtained from
\begin{eqnarray}
{\cal F}_\lambda &=& - k_{\mathrm{B}} T~\lim_{n \rightarrow 0}
\frac{\overline{{\mathcal Z}^{n}} - 1}{n} = \nonumber\\
&=& {\ul12} \int\!\!\int {\mathrm{d}}\vct r {\mathrm{d}} \vct r' \rho_{0}(\vct r) v(\vct r,
\vct r') \rho_{0}(\vct r') + {\ul12} {\rm Tr}~ g(\vct r) v(\vct r, \vct r') - k_{\mathrm{B}} T~\lambda \int\!\! {\mathrm{d}} \vct
R~\Omega (\vct R)~ e^{\beta u(\vct R)}.
\end{eqnarray}
If we now go back from the grand-canonical to the canonical partition function
\cite{netz-SC} again {\em via} a Legendre transform as given in Eqs. (\ref{legendre-F_N}) and 
(\ref{legendre-N-lambda}),  we finally remain with
\begin{equation}
{\cal F}_N = {\ul12} \int\!\!\int {\mathrm{d}}\vct r {\mathrm{d}} \vct r' \rho_{0}(\vct r) v(\vct r,
\vct r') \rho_{0}(\vct r') + {\ul12} {\rm Tr}~ g(\vct r) v(\vct r, \vct r') -
N  k_{\mathrm{B}} T \ln{\!\!\int\!\! {\mathrm{d}} \vct R~\Omega (\vct R)~ e^{\beta u(\vct R)}}.
\label{fin-1}
\end{equation}
The three terms in the above free energy have the following interpretation: the first one is the direct Coulomb interaction
between the fixed charged surfaces (macroions)
 of mean charge density $\rho_0$. The second one, very similar to an analogous term in the 
weak coupling limit, represents the direct effect of the disorder and is proportional to the mean-square disorder charge density $g$. The
third term embodies the disorder effect on the strong coupling level: It depends in an exponential fashion on the mean-square 
disorder strength $g$ and has a non-trivial dependence on Coulomb interaction potential $v(\vct r, \vct r')$ (see Eq. (\ref{u_final})).

\section{Strong-coupling results for two charged plates}

\begin{figure}[t]
\centerline{\includegraphics[height=7cm]{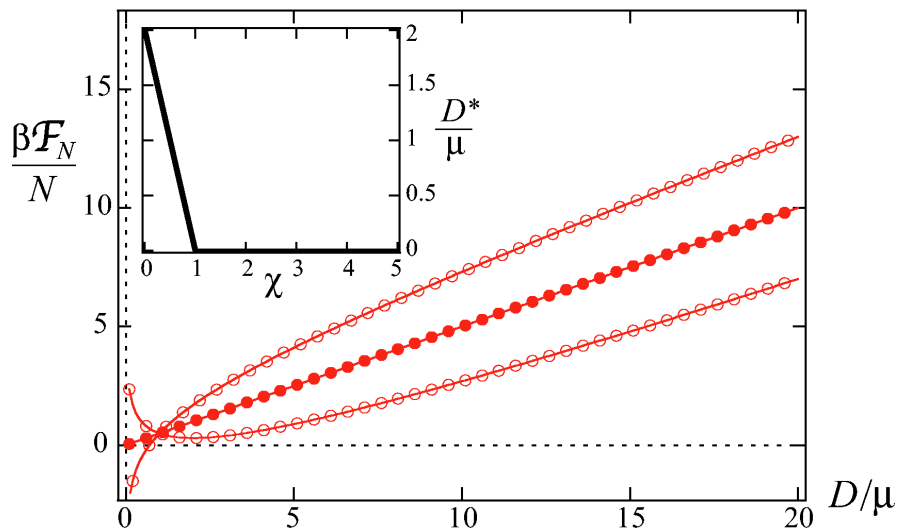}}
\caption{Free energy per particle, Eq. (\ref{FEN-1}), as a function of
dimensionless inter-plate distance $D/\mu$ for $\chi = 0,1$ and 2 (bottom line, middle line, top line). 
 $\mu= e_0/(2 \pi \ell_{\mathrm{B}} Z \sigma)$  is the Gouy-Chapman length for mean surface
charge density. Inset: dimensionless equilibrium distance 
$D^{*}/\mu$ as a function of the disorder coupling parameter $\chi$.}
\label{fig1}
\end{figure}

Assuming again that our system is composed of two planar surfaces, located at $z = \pm a$, and 
with surface charge density $-\sigma$, we have 
\begin{equation}
\rho_{0}(\vct r) = -\sigma \delta(z - a) -\sigma \delta(z + a),  
\end{equation}
where because of the electroneutrality
\begin{equation}
2 S \sigma = N Z e_{0}
\end{equation}
(with $S$ being the area for each surface). The strength of the disorder is assumed to be of  the form 
\begin{equation}
g(\vct r) = g \delta(z - a) + g \delta(z + a),
\end{equation}
and the geometry function $\Omega(\vct R)$ is given by Eq. (\ref{eq:omega}). 
We can now compute all the terms in the expression for the free energy, Eq. (\ref{fin-1}), explicitly.
We obtain
\begin{equation}
{\ul12} \int\!\!\int {\mathrm{d}}\vct r {\mathrm{d}} \vct r' \rho_{0}(\vct r) v(\vct r,
\vct r') \rho_{0}(\vct r') = -{\ul12} \frac{\sigma^{2} S
(2a)}{\epsilon\epsilon_{0}} + const.
\end{equation}
where the term $const.$ again stands for all the terms that are independent of the separation between the
surfaces. Furthermore 
\begin{eqnarray}
  \beta u(\vct R) &=& \beta Z e_{0} \int {\mathrm{d}} \vct r'
v(\vct r', \vct R) \rho_{0}(\vct r') + {\ul12} (\beta Z e_{0})^{2}\int {\mathrm{d}}
\vct r' ~g(\vct r') v^{2}(\vct r', \vct R) = \nonumber\\
&=& - \beta Z e_{0}
\frac{\sigma (2a)}{2 \epsilon\epsilon_{0}} - {\ul12} \frac{(\beta  Z e_{0})^{2} 2
\pi g}{(4 \pi
\epsilon\epsilon_{0})^{2}} \ln{(a^{2} - z^{2})} + const.
\end{eqnarray}
The last term is again independent of the separation between the surfaces and
contributes an additive constant to the free energy. Inserting all these
expressions back into Eq. (\ref{fin-1}), we obtain
\begin{eqnarray}
\frac{\beta {\cal F}_N}{N} &=& {\ul12} \frac{\beta Z e_{0} \sigma}{2
\epsilon\epsilon_{0}} D + (\chi -1) \ln{D} - I(\chi)  \nonumber\\
&=& {\ul12} \frac{D}{\mu} + (\chi -1) \ln{D} - I(\chi)
\label{FEN-1}
\end{eqnarray}
where $D = 2a$ is the separation between the surfaces, $\mu= e_0/(2 \pi \ell_{\mathrm{B}} Z \sigma)$ 
is just the Gouy-Chapman length for mean surface charge
density, and
\begin{equation}
\chi = \frac{Z^{2} (\beta e_{0})^{2} g}{8 \pi (\epsilon\epsilon_{0})^{2}} = 2\pi Z^2\ell_{\mathrm{B}}^2 g
\end{equation}
is the dimensionless  {\em disorder coupling parameter}. Note that it is very 
similar to the electrostatic coupling parameter for mean charge $\Xi$, Eq. (\ref{xi}), except that
it scales with counterion valency as $Z^{2}$ instead of $Z^{3}$. Furthermore
\begin{equation}
I(\chi) = \ln{\frac{1}{D} \int_{-D/2}^{+D/2} \left( \frac{(D/2)^2- z^{2} }{D^{2}}\right)^{-\chi/2} {\mathrm{d}}z} = ~_{2}F_{1}(\ul12,
{\textstyle{\frac{\chi}{2}}}, \ul32, 1),
\end{equation}
which is independent of the inter-plate distance $D$. 
The effect of disorder on the interactions between disordered charged  surfaces in the
strong-coupling limit is thus contained in the second term in Eq. (\ref{FEN-1}). Its behavior
depends crucially on whether the dimensionless disorder  coupling strength is
smaller, equal or larger than one. Evaluating the interaction pressure, $P(D)$,  from
the free energy, Eq. (\ref{FEN-1}), we find, in dimensionless representation, 
\begin{equation}
\tilde P(D) \equiv  \frac{\beta P(D)}{2\pi \ell_{\mathrm{B}}\sigma^2} = \tilde P_{\mathrm{SC}}(D) - {\chi}\bigg(\frac{2\mu}{D}\bigg),
\end{equation}
where the first term is the standard strong-coupling pressure \cite{netz-SC}
\begin{equation}
 \tilde P_{\mathrm{SC}}(D) =   -1+ \frac{2\mu}{D}, 
\end{equation}
while the second additive term represents the effects of the
disorder (see Fig. \ref{fig1}).

From Eq. (\ref{FEN-1}), we can derive also the equilibrium distance $D^{*}$ between
the two surfaces, corresponding to zero interaction pressure. It is obtained as 
\begin{equation}
D^{*} =  2~ (1 - \chi)~\mu.
\end{equation}
In the undisordered case, $\chi = 0$, this reduces to the known form as obtained in Ref. 
\cite{netz-SC}, that is an equilibrium distance equal to twice the Gouy-Chapman length. From this value the
equilibrium distance then continuously approaches towards zero for increasing  $\chi$; it reaches zero for
the critical value $\chi_c = 1$ and remains at zero thereafter. This behavior has all the features of a second-order, quenched-disorder-driven
collapse transition with an unusual value of the critical exponent (see Fig. \ref{fig1}, inset).

For $\chi = 1$ the interaction pressure between the surfaces is obviously constant in the whole
range of separations $D$ right down to zero. For $\chi \geq 1$, the counterion entropy, {\em i.e.} the $- \ln{D}$ term
in Eq. (\ref{FEN-1}), is completely wiped out by the surface charge disorder. It thus appears that the quenched charge
disorder and the counterion entropy for large values of the disorder coupling strength in some sense counteract one
another.

\section{Discussion}

We have assessed the effect of the quenched macroion charge disorder in the case of 
interactions between two apposed surfaces in a one-component Coulomb fluid. We assumed a Gaussian {\em ansatz} 
for the quenched surface charge  density distribution, with finite mean and a constant
mean-square width. Though the final results within these assumptions are very
simple they are highly non-trivial and difficult to derive. 

In the case of  two ideally polarizable surfaces
at separation $D$, we find that in the weak-coupling or the Poisson-Boltzmann level, $\Xi\ll 1$ (where 
$\Xi=2\pi Z^3 \ell_{\mathrm{B}}^2 \sigma$ is the electrostatic coupling due to the mean charge $\sigma$), the quenched
surface charge disorder has no effect on the interactions. This is strictly
true only  for ideally polarizable surfaces. In the general case, electrostatic
images would contribute to the disorder-driven part of the interaction, making
it non-zero and dependent on the dielectric mismatch between the dielectric
material of the surfaces and that of the aqueous solution bathing the 
electrolyte in between. The disorder-driven part of the interaction in this general case can be
either attractive or repulsive, depending on the sign of the dielectric
mismatch  \cite{paper-2}.

On the strong-coupling level $\Xi\gg 1$, the disorder effects are always present even in
the case of ideally polarizable bounding surfaces. The disorder-driven part of
the interaction free energy turns out to be additive and is linear in the
disorder coupling parameter $\chi= 2\pi Z^2 \ell_{\mathrm{B}}^2 g$, which is analogous to the mean charge coupling
parameter $\Xi$ introduced in Ref. \cite{netz-SC}, and scales with the square counterion valency $Z$ and 
 the mean-square width of the disorder distribution $g$. 
For increasing  disorder coupling parameter
$\chi$, we find that the equilibrium distance is gradually displaced towards smaller values
if compared to the case with no disorder, $\chi = 0$. At a critical value
$\chi_{c} = 1$, the equilibrium distance becomes zero and remains zero for even
larger values of the disorder coupling parameter. Thus one could think of the
disorder effects in terms of a disorder-driven second-order collapse phase
transition.
We are unsure at this point how
general this interpretation might be. In order to assess its validity
additional cases of disorder modified Coulomb interactions should be
investigated. 

Since quenched disorder is ubiquitous in the context of charged biological macroions, our analysis should be of
fundamental value especially in this context. In understanding the electrostatic interactions between charged
membranes, polyelectrolytes and cells, sufficient attention should thus be given to disorder effects that obviously
contribute in a non-trivial manner to the electrostatic interactions in aqueous electrolyte solutions. Pronounced
attraction stemming from the strong-coupling electrostatics can thus be made even stronger due to charge disorder
effects. 

From our general formulation it follows that even for net-neutral macroions, {\em
i.e.} $\sigma = 0$, but with
non-zero charge disorder, $g \neq 0$, the total interaction can show electrostatic effects! This is indeed the most
non-intuitive and striking feature of our theory. Persistent observations of
electrophoretic potentials \cite{elefor} of net uncharged
particles might thus have an elegant explanation in terms of the disorder effects invoked in this contribution.
These might enhance the effects of other charging mechanisms such as Lifshitz-van der Waals-interactions driven 
adsorption or might in fact, if proper source of the quenched disorder is identified, be the sole cause of electrostatic 
effects with nominally uncharged   macroions. Concurrent measurements of the electrophoretic potentials with an estimation of
the local macroion charge disorder could substantiate or disprove the explanation tentatively proposed above.

Numerical simulations of counterions in the presence of disordered surface charge distributions
will be valuable to examine the present predictions, in particular, for large mean electrostatic coupling $\Xi\gg 1$ 
and large disorder strength, where  the disorder-driven collapse transition is predicted. 
Simulations with disorder effects are however difficult and time consuming and we leave them for a subsequent 
publication \cite{paper-2}. We have also not given appropriate attention to the image effects in this contribution. They 
would add additional details to the disorder-driven interaction that would tend to obscure our main points. For reasons 
of clarity we will thus relegate the image effects to a subsequent publication \cite{paper-2}.

\section{Acknowlegements}

RP would like to acknowledge financial support of the Slovenian Research Agency under Grant No. P1-0055.
AN acknowledges financial support from the DFG German-French Network. 



\end{document}